\documentclass[amssymb,amsmath,nofootinbib]{revtex4}

\usepackage[all]{xy}
\usepackage{color}
\usepackage{subfigure}
\usepackage{bm,amsmath}
\usepackage{marvosym}
\usepackage{graphicx}
\usepackage{color}
\usepackage{wrapfig}
\usepackage{epsf}
\usepackage{times,mathptm}
\usepackage{url}
\usepackage{clrscode}

\def\(({\left(}
\def\)){\right)}
\def\[[{\left[}
\def\]]{\right]}

\newcommand{\be}{\begin{equation}}
\newcommand{\ee}{\end{equation}}
\newcommand{\bea}{\begin{eqnarray}}
\newcommand{\eea}{\end{eqnarray}}

\begin{document}

\title{Message Passing for Optimization and Control of Power Grid:\\ {\it Model of Distribution System with Redundancy}}
\author {Lenka Zdeborov\'a$^1$, Aur\'elien Decelle$^{1,2}$, and Michael Chertkov$^1$}

\affiliation{$^1$ Center for Nonlinear Studies and Theoretical Division, LANL, Los Alamos, NM 87545 \\
$^2$  Universit\'e Paris-Sud \& CNRS, LPTMS, UMR8626,  B\^{a}t.~100, Universit\'e
Paris-Sud 91405 Orsay cedex}

\date{\today}

\begin{abstract}
We use a power grid model with $M$ generators and $N$ consumption units to optimize the grid and its control. Each consumer demand is drawn from a predefined finite-size-support distribution, thus simulating the instantaneous load fluctuations. Each generator has a maximum power capability. A generator is not overloaded if the sum of the loads of consumers connected to a generator does not exceed its maximum production.  In the standard grid each consumer is connected only to its designated generator, while we consider a more general organization of the grid allowing each consumer to select one generator depending on the load from a pre-defined consumer-dependent and sufficiently small set of generators which can all serve the load. The model grid is interconnected in a graph with loops, drawn from an ensemble of random bipartite graphs, while each allowed configuration of loaded links represent a set of graph covering trees. Losses, the reactive character of the grid and the transmission-level connections between generators (and many other details relevant to realistic power grid) are ignored in this proof-of-principles study. We focus on the asymptotic limit, $N\to\infty$ and $N/M\to D=O(1)>1$, and we show that the interconnects allow significant expansion of the parameter domains for which the probability of a generator overload is asymptotically zero. Our construction explores the formal relation between the problem of grid optimization and the modern theory of sparse graphical models. We also design heuristic algorithms that achieve the asymptotically optimal selection of loaded links.  We conclude discussing the ability of this approach to include other effects, such as a more realistic modeling of the power grid and related optimization and control algorithms.
\end{abstract}

% LA-UR 09-02066

\maketitle

\section{Introduction}

 The existing power grid is complex and far from being optimized. The anticipated installation of small-scale distributed generators and storage devices, as well as the addition of many ancillary backup lines and control devices at both transmission and  distribution levels, imply that the main future challenges will require intelligent planning, optimization and control of this ever growing grid. This optimized and efficient grid of the future, that incorporates new hardware and concepts such as renewable generation and distributed storage, has been labeled "smart grid" \cite{08DOE,09SG}. Accounting for many important details of the power distribution and transmission is not feasible without much simplification. Simplified models must identify the significant effects, extracting and analyzing and later probing each of them  separately and in combinations. (See, e.g. related review articles \cite{05AW,07Ili,08Ami}.)

We adopt this "discovery though simplification" approach and focus in this paper on improving the functioning and control of the electric grid on the power distribution level.  Specifically, our prime focus is on preventing overloads of the power generation units caused by fluctuations in the demand by efficiently utilizing ancillary lines. This approach is justified in the context of a city-scale transmission system, in which many significant loads and generators are in geometric proximity of each other and the cost of building new ancillary lines is not prohibitive. We assume that shedding excess load is not an option and we consider the possibility of redistributing the load via a system of interconnects over a larger grid than what is used under normal (no overload) conditions. Fundamentally,  we ask the following question: can an intelligent arrangement of $O(M)$ ancillary lines among a system of $M$ generators, each serving $D$ customers,  possibly reduce the probability of a generator failure from a finite number to zero in the limit of $M\to\infty$  and $D=O(1)$?  In fact,  an assumption that a drastic improvement is possible stems from basic information-theoretic intuition: any finite error probability can be reduced to zero via properly introduced redundancy \cite{48Sha}. We show that adding interconnects to the power system is related to adding redundancy to the information system, in the sense that achieving the asymptotic overload-free distribution is indeed a possibility for an idealized grid.

The model of power line considered in the paper mimics how the ancillary power lines are operated on the distribution level power grid. A city-scale power grid has an intervene loopy structure,  however these loops are typically used for ancillary (backup and/or maintenance) purposes and operators aim at avoiding running current over loops.   We adopt this strategy and assume that the system contain (or may be built with) many switches and that it is operated in the manner that
for each given configuration of the pre-installed on-off switches the currents flow over trees, i.e. subgraphs of the full loopy distribution graph without loops. We also assume that the lines are sufficiently short (kilometers, not hundreds or thousand of kilometers), and thus thermal losses are not important and reactive parts of line impedances can be ignored. These assumptions correspond to the so-called DC approximation \cite{96WW}. Combination of the loop-free structure of power flows (for any given configuration of switches) with DC approximation allows to model electricity delivery as an abstract commodity flow \cite{90CLR}. For completeness of our description,  let us also note that some other effects of realistic power flows are ignored in our first publication on this emerging subject.
Thus, we do not consider back flows (consumers turned into distributed generators) and for that matter we do not discuss at all the entire scale up (transmission level) structure of the grid,  inhomogeneities and spatio-temporal correlations in loads and generation, and all economy, pricing and regulation effects \cite{01BH}. In essence, our main task here consists in establishing the existence of fundamental limits, and bounds on the idealized asymptotic failure-free regime, and followed by developing efficient and simple algorithm controlling switching in the grid.

To solve our model we use the cavity method  and its computational realization via population dynamics, introduced in the statistical physics of disordered systems \cite{85MP,01MP} and recently adapted to the analysis of Shannon (phase) transitions in constraint satisfaction \cite{02MPZ} and error-correction \cite{08RU,09MM}. This method explores the famous fact that the Bethe-Peierls \cite{35Bet,36Pei} or Belief Propagation \cite{63Gal,88Pea} (BP) scheme exactly solves probabilistic models on graphs without loops (a tree). The method allows to evaluate the ensemble average over configurations of allowed discrete switchings on the grid balancing load with generation. We study how the number of allowed solutions scale with the size of the grid, and identify a transition (in the space of model parameters) from a satisfiable (SAT) domain, in which the number of solutions is exponentially large (in the system size) and where BP algorithm finds a valid solution easily, to an unsatisfiable (UNSAT) domain in which the number of valid solutions is with high probability zero (thus load shedding would be required).

Assuming that the two ways communications between consumers and generators exists we design a stochastic local search algorithm, coined WalkGrid, that is able to find optimal configuration of switches almost anywhere in the SAT phase. We also utilize aforementioned BP analysis and develop a BP-based message passing scheme for efficient search of SAT configuration of switches. Generally,  WalkGrid outperforms the BP-based algorithm almost anywhere in the SAT phase.  Note however that one useful feature of BP (not readily available in WalkGrid) consists in its ability to count number of available SAT solutions, and thus have a direct algorithmic test of the distance to failure,  i.e. distance to the SAT-UNSAT transition.

The material in the manuscript is organized as follows. We introduce our model of the distribution grid in Section \ref{sec:Toy}. The Belief Propagation method is detailed in Section \ref{sec:BP}. We describe population dynamics algorithms and present SAT-UNSAT transition results in \ref{sec:PD}. Control algorithms are explained  in Section \ref{sec:Control}. Section \ref{sec:Con} summarizes our approach and proposes extensions of the study for future explorations relevant to intelligent optimization and control of the power grid.

\section{Grid model}
\label{sec:Toy}

Consider $M$ sources/generators each connected to $D$ distinct consumers, so that the total number of consumers is $M D$. Greek/Latin indices will be reserved for generators/consumers. For simplicity we assume that each generator has a maximum production rate of unity, i.e. ${\bm y}=(0\leq y_\alpha\leq 1|\alpha=1,\cdots,M)$, though inhomogeneities in the production can be easily incorporated into our approach.
The configuration of loads, ${\bm x}=(x_i>0|i=1,\cdots,N)$, is drawn from an assumed known distribution with support on the interval $(0,1)$. Our enabling example is the flat-box ensemble with a fraction $\epsilon$ of customers dropped off from the network, i.e. drawing no electricity at all,
\begin{eqnarray}
{\cal P}({\bm x})=\prod_i [ (1-\epsilon) p(x_i) + \epsilon \delta(x_i)],\quad p(x)=\left\{
\begin{array}{cc}
1/\Delta, & |x-\bar{x}|<\Delta/2\\
0, & \mbox{otherwise}
\end{array}
\right. \label{Px}
\end{eqnarray}
The mean consumption then is $(1-\epsilon)\overline x$, and $\Delta$ is the width of the part of the distribution,  correspondent to nonzero demand. We require that the consumption of each individual customer is non-negative \footnote{This assumption is not crucial and can be easily removed, allowing consumers not only to consume but also to generate electricity. The inclusion of this capability potentially has a far reaching consequences because it allows distributed generation. We postpone our discussion of this interesting possibility to future publications.} , i.e. \be 2\bar{x} \ge \Delta, \label{cond_1}\ee and we
assume that under normal conditions, i.e. when demand is not excessive,  there is always enough power produced in the network to supply electricity to all consumers, i.e.
\be (1-\epsilon)\bar{x}\le \frac{M}{N} = \frac{1}{D}\, . \label{cond_2}
\ee

In our "standard" model of the grid each consumer is assigned to strictly one generator, while  each generator feeds exactly $D$ consumers. This corresponds to a graph decomposed into $M$ simple trees, where $M$ is the number of generators.  To guarantee in this standard case that a generator is always SAT, i.e. it is capable of supplying electricity to all of its customers, one requires
\be
        D\left(\bar{x}+\frac{\Delta}{2}\right) \le 1 \, .\label{separated}
\ee
This later condition will be challenged below, in the sense that we will show that by adding ancillary lines one does not need to impose (\ref{separated}), but instead have a weaker condition, thus extending the domain of the asymptotically failure-free generation.

\begin{figure}
\centering
\subfigure[\hspace{0.3cm} $R=0,1,2,3$. Graph samples. Ancillary connections to foreign generators/consumers are shown in color.]{\includegraphics[width=0.4\textwidth,page=1]{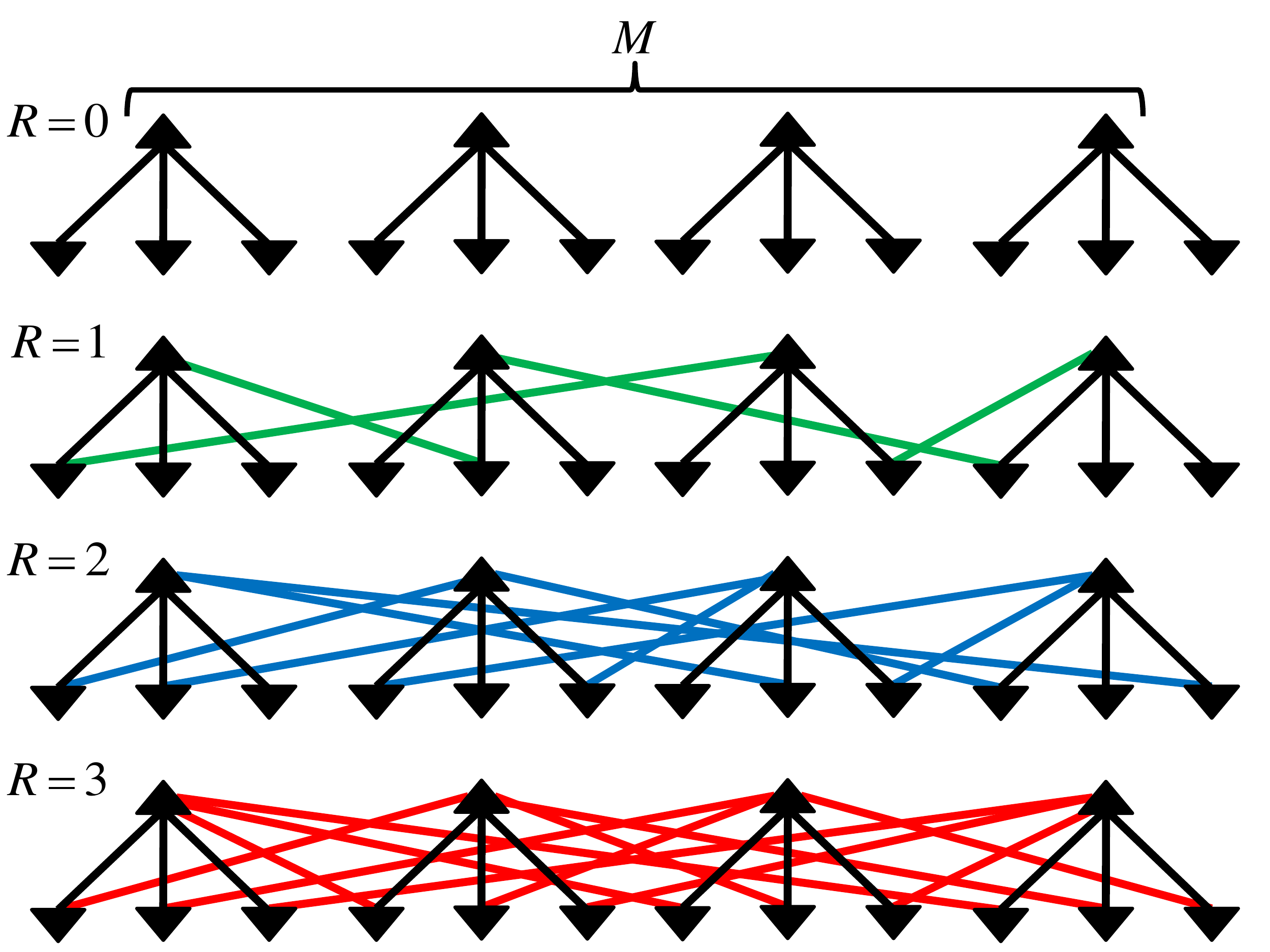}}
\subfigure[\hspace{0.3cm} $R=1$. Three valid (SAT) configurations (shown in black, the rest is in gray) for a sample graph shown in Fig.~\ref{graphs}a.  ]{\includegraphics[width=0.4\textwidth,page=2]{graphs.pdf}}

\caption{ (Color online) Illustration of the bipartite graph construction of our grid with $D=3$.  \label{graphs} }
\end{figure}

To achieve this improvement we consider an interconnected grid built from the standard/separated one by adding ancillary lines between consumers and generators. For our quantitative analysis we choose a random graph drawn  in the following way. First, $D$ consumers are connected to every producer. Second, we choose $R$ consumers from each producer and connect them to a second producer in such a way that every producer is connected to $D+R$ consumers. Of these $D+R$ connections $D-R$ go to singly connected consumers and $2R$ to doubly connected consumers. So that a total of $NR/D$ consumers are connected to two generators and $N(1-R/D)$ are connected to one generator. See Fig.~\ref{graphs} for illustration. The choice of the two parameter $(R,D)$ graph ensemble is used for simplicity. Our analysis method is valid for any random network as long as it is bipartite and locally tree-like, meaning that the length of the shortest loop going through a random node is $O(\log{N})$.

We say that {\it given configuration of loads} ${\bm x}=(x_i|i=1,\cdots,N)$ {\cal is SAT} (satisfiable) if there exists a matching ${\bm \sigma}=(\sigma_{i\alpha}=0,1|\{i,\alpha\}\in{\cal G})$, where ${\cal G}$ is the bi-partite graph accounting for all generators, consumers and lines, such that the following set of conditions are simultaneously satisfied
\begin{eqnarray}
&& \forall i\in{\cal G}:\quad \sum_{\alpha\in\partial i}\sigma_{i\alpha}=1,\label{mat_cond}\\
&& \forall \alpha\in{\cal G}:\quad \sum_{i\in\partial \alpha}\sigma_{i\alpha} x_i\leq 1\, , \label{cap_cond}
\end{eqnarray}
where $\partial i$, resp. $\partial \alpha$, stand for all the nodes to which $i$, resp. $a$, is connected.
If the reverse is true, i.e. there exists no valid ${\bm \sigma}$ with all Eqs.~(\ref{mat_cond},\ref{cap_cond}) satisfied, we say that ${\bm x}$ is UNSAT (unsatisfiable). First, we aim to solve the decision problem: is ${\bm x}$ is SAT or UNSAT?
Furthermore, if ${\bm x}$ is SAT, we would like to find at least one valid solution, ${\bm\sigma}$. Both problems can be stated for a given
graph or, alternatively,  can be considered "in average" for ensemble of graphs. We describe the "average" solution of the former problem in Sections \ref{sec:PD} and the algorithmic solution of the latter in Section \ref{sec:Control}. Both Sections will be preceded by a preparatory discussion of the BP approach in Section \ref{sec:BP}.

Related models have been studied in the context of resource allocation. One example is the problem of online advertising known as {\it AdWords}, in particular its uniform off-line version  \cite{01LLN,04AM,08ABKMN}. The main difference between our model and the model for budget-constrained advertising problem is that condition (\ref{cap_cond}) is replaced by the maximization of the total revenue $\sum_{\alpha} \min [ 1, \sum_{i\in \partial \alpha} \sigma_{i\alpha} x_i ]$. This relaxation translates into allowing one to shed loads, that is not an option in our setting. Discussion of the BP-based approaches to the AdWords problem and their methodology in \cite{09ABRZ} is very much related to ours.

Of other models developed in Computer Sciences,  the one most similar to ours is the so-called off-line weighted balls-into-bins games (sometimes called balanced loading) \cite{TalwarWieder07,BerenbrinkFriedetzky08}. The off-line version of the balls-into-bins games is known to be polynomial in the non-weighted case (i.e. for a special discrete choice of the distribution (\ref{Px})), as in that case it can be mapped into a max-flow (min-cut) problem \cite{SandersEgner03}. On the other hand the general (weighted) version on a fully connected graph (every customer being possibly connected to every generator) is equivalent to the NP-complete bin packing problem (number partitioning in the case of two generators). We are not aware of any computational complexity results for the case of bounded number of consumers per generator. Therefore, and given that the max-flow mapping of  \cite{SandersEgner03} does not generalize to the most general version of Eq.~(\ref{Px}), we conjecture that computationally the general problem described by Eq.~(\ref{Px}) is NP-hard.

\section{Belief Propagation}
\label{sec:BP}

In the asymptotic limit of an infinite system, for which $N\to \infty$ while  $D$ and $R$ are $O(1)$, the interconnected grid is locally tree-like. Therefore, the Bethe-Peierls (Belief Propagation) BP approach for evaluating the generalization of (\ref{separated}) is expected to be asymptotically exact (in a sense to be clarified later in this text). This Section describes details of the BP approach

We introduce the following set of {\it marginal probabilities}:
\begin{itemize}
 \item $\psi_1^{\alpha\to i}$ is the probability that generator $\alpha$ is satisfied given that the edge $(i,\alpha)$, connecting $\alpha$ with his consumer-neighbor on ${\cal G}$, is in the active state, i.e. $\sigma_{i\alpha}=1$.

 \item $\psi_0^{\alpha\to i}$ is the probability that generator $\alpha$ is satisfied given that the edge $(i,\alpha)$, where
 $(i,\alpha)\in{\cal G}_1$, is inactive, i.e. $\sigma_{i\alpha}=0$.

 \item $\chi_1^{i\to \alpha}$ is the probability that $i$ is satisfied (i.e. it is connected to exactly one generator) given that
 the edge $(i,\alpha)$, where
 $(i,\alpha)\in{\cal G}_1$, is active, i.e. $\sigma_{i\alpha}=1$.

 \item $\chi_0^{i\to \alpha}$ is the probability that $i$ is satisfied (i.e. it is connected to exactly one generator) given that
 the edge $(i,\alpha)$, where
 $(i,\alpha)\in{\cal G}_1$, is inactive, i.e. $\sigma_{i\alpha}=0$.
 \end{itemize}
BP  relates these marginal probabilities to each other assuming that the relations are graph local, i.e. as if the graph would  contain no loops. The resulting Belief Propagation (BP) equations are
\bea \label{BP}
          \chi_1^{i\to \alpha} &=& \frac{1}{Z^{i\to \alpha}} \prod_{\beta\in \partial i \setminus \alpha} \psi_0^{\beta\to i}\, , \label{BP1}\\
         \chi_0^{i\to \alpha} &=& \frac{1}{Z^{i\to \alpha}} \sum_{\beta\in \partial i \setminus \alpha} \psi_1^{\beta\to i}
         \prod_{\gamma\in \partial i \setminus \alpha,\beta} \psi_0^{\gamma\to i}  \, , \label{BP2}\\
          \psi_1^{\alpha\to i} &=&  \frac{1}{Z^{\alpha\to i}}  \sum_{{\bm\sigma}_{\partial\alpha\setminus i\alpha}=\{0,1\}^{|\partial_\alpha\setminus i|}}
          \theta(1-x_{i}-\sum_{j\in \partial \alpha \setminus i} \sigma_{j\alpha} x_{j}) \prod_{j\in \partial \alpha \setminus i} \chi_{\sigma_{j\alpha}}^{j\to \alpha}\, , \label{BP3}\\
 \psi_0^{\alpha\to i} &=&  \frac{1}{Z^{\alpha\to i}}  \sum_{{\bm\sigma}_{\partial\alpha\setminus i\alpha}=\{0,1\}^{|\partial_\alpha\setminus i|}}
  \theta(1-\sum_{j\in \partial \alpha \setminus i} \sigma_{j\alpha} x_{j}) \prod_{j\in \partial \alpha \setminus i} \chi_{\sigma_{j\alpha}}^{j\to \alpha} \, ,\label{BP4}
\eea
where  $\partial i\setminus\alpha$ is a standard notation for the set of generator nodes linked to consumer $i$, however excluding generator $\alpha$, and similarly $\partial\alpha\setminus i$ stands for the set of consumer nodes linked to generator $\alpha$ excluding consumer $i$. In Eqs.~(\ref{BP}-\ref{BP4}), $Z^{i\to \alpha}$ and $Z^{\alpha\to i}$ are normalizations ensuring that, $\chi^{i\to \alpha}_1+\chi^{i\to \alpha}_0=1$ and
$ \psi_1^{\alpha\to i} + \psi_0^{\alpha\to i} =1$. The $\theta(\cdot)$ is the step function enforcing the generation constraints. It is unity if the argument is positive and zero otherwise. The probability for the link/edge $(i,\alpha)$ to be active, stated in terms of the related $\psi$ and $\chi$, is
\be
       p(i,\alpha)= \frac{\psi_1^{\alpha\to i}\chi_1^{i\to \alpha}}{\psi_1^{\alpha\to i}\chi_1^{i\to \alpha}+
       \psi_0^{\alpha\to i}\chi_0^{i\to \alpha}}\, ,\label{marg}
\ee
where $\sum_{\alpha \in \partial i} p(i,\alpha)=1$. Note, for the sake of completeness,  that the BP equations can be derived in the spirit of \cite{01YFW} as conditions for a fixed-point (minimum) of a functional of marginal probabilities (beliefs), the so-called Bethe free energy functional.

The Bethe entropy, defined as the logarithm of the number of possible SAT configurations (and also equal to the Bethe free energy evaluated at the solution of the previously mentioned BP equations), is
\bea
   S_{\rm Bethe} &=& \sum_{\alpha} \log(Z^{\alpha})  + \sum_{i} \log(Z^{i})  - \sum_{(i,\alpha)} \log(Z^{i\alpha})\, , \\
   Z^{\alpha} &=&  \sum_{{\bm\sigma}_{\partial\alpha}=\{0,1\}^{|\partial_\alpha|}}
    \theta(1-\sum_{i\in \partial \alpha} \sigma_{i\alpha} x_{i}) \prod_{i\in \partial \alpha} \chi_{\sigma_{i\alpha}}^{i\to \alpha}  \, , \\
Z^{i}&=&  \sum_{\alpha\in \partial i } \psi_1^{\alpha\to i}  \prod_{\beta\in \partial i \setminus \alpha} \psi_0^{\beta\to i}  \, ,  \\
Z^{i\alpha} &=&\psi_1^{\alpha\to i}\chi_1^{i\to \alpha}+\psi_0^{\alpha\to i}\chi_0^{i\to \alpha}\, .
\label{entr}
\eea
The entropy $S_{\rm Bethe}$ is extensive, $O(N)$, and self-averaging, i.e. the distribution of $S_{\rm Bethe}$ is concentrated around its mean value with dispersion being $o(N)$ at $N\to\infty$.

\section{Average Bethe Entropy via Population Dynamics}
\label{sec:PD}

Fixed point of the BP Eqs.~(\ref{BP}-\ref{BP4}) and the corresponding entropy (\ref{entr}) can be obtained by solving Eqs.~(\ref{BP}-\ref{BP4}) iteratively for a given instance of the problem. Repeating these simulations many times for different instances of the graph and load ensembles one can also calculate the average Bethe entropy. However, the average behavior, corresponding to the limits of the infinite graph, can be obtained more efficiently via the population dynamics technique. This technique offers a computationally efficient sampling from the distribution of the marginal probabilities $\chi_{\sigma}$ on infinite random graphs and subsequent evaluation of the average Bethe entropy.  Below we give a very brief exposition of the population dynamics approach applied to our model. The interested reader is referred to \cite{01MP,08Zde} for further details.

In population dynamics, we create a pool of $N_p$ components, each characterized by the $(\chi_0,\chi_1,x)$ vector with $x$ drawn from the original distribution of demands (\ref{Px}) and the initial $\chi$ selected arbitrarily.  Messages leaving the generator and entering the consumer connected to a single generator will always be kept fixed to  $\chi_1=1$, $\chi_0=0$, while other messages will be updated iteratively, so that at any new sweep a new pool is derived from the old one. A sweep consists of the following step repeated $O(N_p)$ times (each time one of the $N_p$ components is updated). A step consists of choosing a random number $x_i$, $D-R$ random numbers $x_j$, and $2R$ random elements from the pool representing the incoming messages $\chi$, and computing the message $\psi$ based on Eqs.~(\ref{BP3},\ref{BP4}). Then we replace a random element of the pool by the vector $(\psi_1,\psi_0,x_i)$, in accordance with  Eqs. (\ref{BP1}-\ref{BP2}), thus guaranteeing that $\chi_1=\psi_0$ and $\chi_0=\psi_1$. We repeat this sweep procedure many times until convergence is achieved. To compute the average Bethe entropy we apply similar procedure of sampling from the resulting pool and thus averaging all the terms in Eq.~(\ref{entr}).

\begin{figure}[!ht]
  \begin{center}
  \resizebox{0.48\linewidth}{!}{\includegraphics{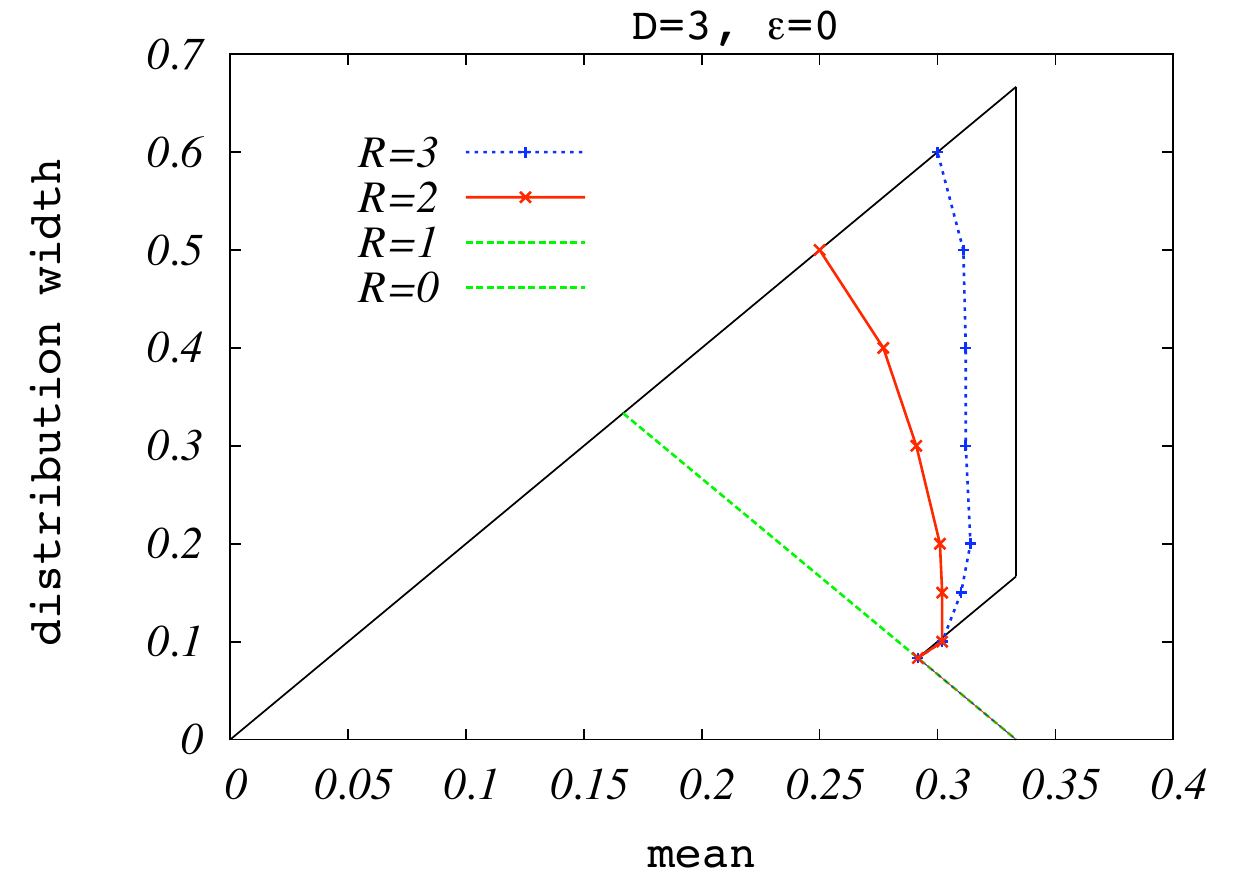}}
  \resizebox{0.48\linewidth}{!}{\includegraphics{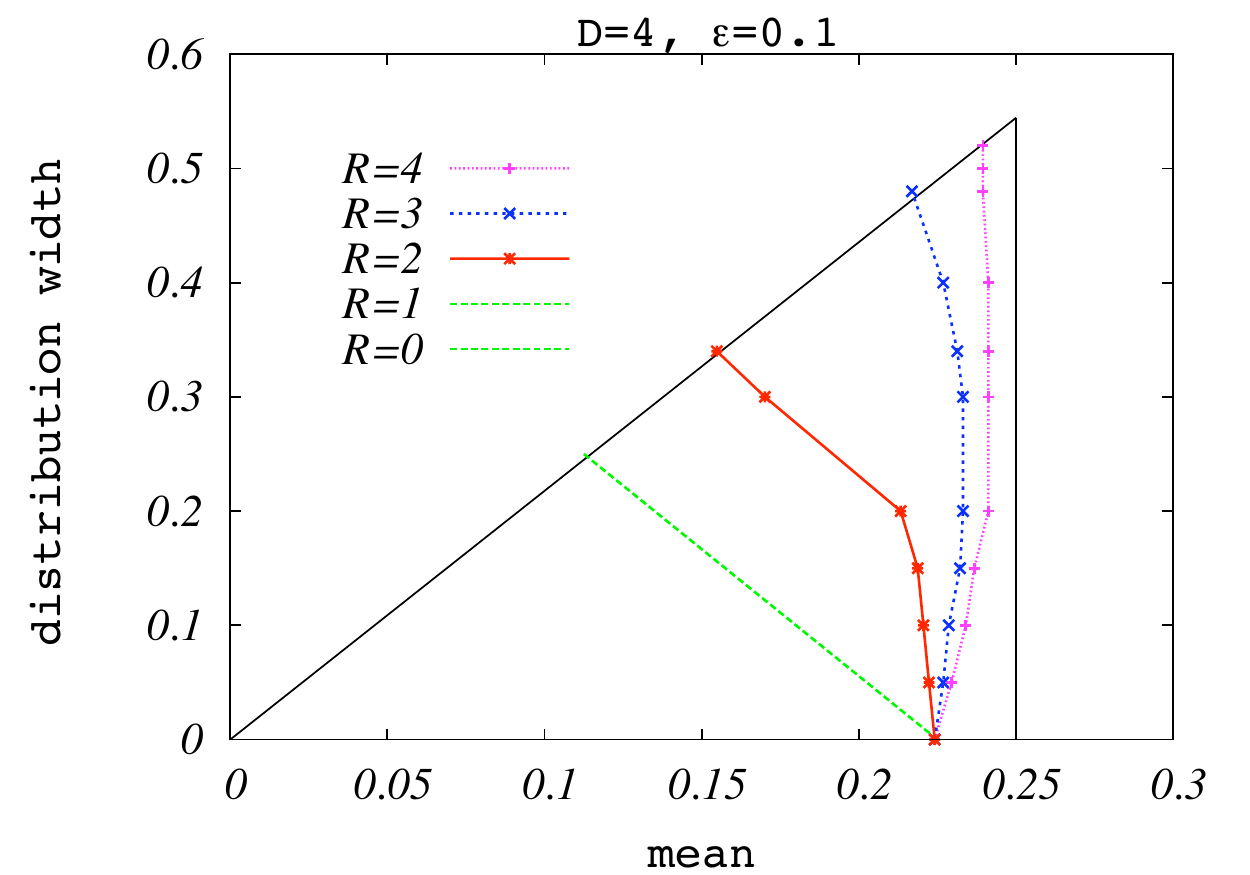}}
  \end{center}
  \caption{ (Color online) 
  Results of the population dynamics shown in the $(\bar{x},\Delta)$ plane for $D=3,4$, where $(1-\epsilon)\bar{x}$ is the distribution mean and $\Delta$ is the distribution widths. The population dynamics was done with $N_P=1000$ for $D=4$ and $N_p=10000$ for $D=3$. The triangular region (black full lines) correspond to conditions (\ref{cond_1}), (\ref{cond_2}) and (\ref{triangle}), and encloses the interesting region of parameters. The colored lines with data points separate SAT and UNSAT domains lying on the bottom-left and top-right off the lines respectively. Different colors/lines/markers correspond to different values of $R=0,\cdots,D$. Note that the curves for $R=0$ and $R=1$ (dashed without markers) are identical, due to condition (\ref{condR}). We observe that the performance improves with increasing $R\ge 2$ (the number of ancillary lines) and the grid tolerates larger values of $\Delta$ (fluctuations in the demand).  The abrupt change in slope of the curve corresponding to $R=2$ (full red line with x marks) in the right part of the figure is due to condition (\ref{condR}). Note that if condition (\ref{cond_1}) is removed (see e.g. discussion in the footnote preceding the equation) our description still remains valid, thus resulting in the colored lines with markers extending smoothly beyond their crossings with the tilted black curve. \label{fig_L2_K6}}
\end{figure}

Implementing the population dynamics method we observed three possible outcomes
\begin{itemize}
   \item[(a)] SAT phase: The Bethe entropy is positive suggesting that the number of SAT configuration (valid redistribution of the demand over the generators) is exponential in the system size \footnote{This is the case unless the so called replica symmetry breaking takes place in our model. We have done a local stability check of the BP solution and have not seen any indication for a break down of the replica symmetry. Note also that this stability is also equivalent to convergence of the underlying BP algorithm for an individual realization of the graph and the loads.}.
   \item[(b)] UNSAT phase, type 1: The Bethe entropy is negative, suggesting that there is almost surely no valid redistribution of demand over the generators.
   \item[(c)] UNSAT phase, type 2: A contradiction is encountered in the BP equations, formally correspondent to zero values for the  normalizations in (\ref{BP}-\ref{BP4}). We conclude that the demand is incompatible with the graph and respective generator assignment. \end{itemize}
Fig.~\ref{fig_L2_K6} shows results of the population dynamics simulations, with lines connecting the marks indicating the boundaries of the respective SAT and UNSAT phases. These boundaries are established by fixing $\Delta$ and traversing different values of $\bar{x}$ starting from SAT phase moving towards UNSAT phase and catching the value where the UNSAT conditions are first time observed. We observe a significant improvement of the SAT/UNSAT threshold from what Eq.~(\ref{separated}) suggests.

Recall that in the  $R=0$ case the SAT/UNSAT threshold is  described by (\ref{separated}). Similar upper bound can be derived for cases $R\ge 1$. Then  for any given $R$, one can find (with high probability) a place on a very large network where all consumers connected to two generators (which are neighbors trough one of the consumers) have demands near to the maximum value $\overline x + \Delta/2$. Then at least $D-R+1$ of these loads have to be connected to one generator and hence
\be
        (D-R+1)(\overline x + \frac{\Delta}{2}) \le 1\, . \label{condR}
\ee
Note that (\ref{condR}) is identical to (\ref{separated}) for $R=1$.

Moreover, to observe an improvement  in the condition (\ref{separated}) at $\epsilon=0$ one needs to require existence of such configuration of loads that a set of $(D+1)$ consumers connected to one producer and drawing the minimal amount of electricity does not overload the generator, i.e. 
\be
     \epsilon=0: \quad (D+1)(\overline x - \frac{\Delta}{2}) \le 1 \, .\label{triangle}
\ee

To conclude, this study of the average Bethe entropy shows that the network with added ancillary lines is able to withstand larger fluctuations in the demand, $\Delta$, than the naive network, in which every consumer has a pre-designed provider independently of the current demand.  This effect is amplified with increasing $R$.

At this point it is also appropriate to recall that the SAT-UNSAT transition is actually an abrupt transition only in the sense of the asymptotic $N\to\infty$ limit. Thus for large finite $N$ the generator failure probability is small but finite at any point of the SAT domain.

\section{Control Algorithms}
\label{sec:Control}

In this Section we discuss the problem of calculating a valid load-to-generators assignment for a given graph and given configuration of loads. We designed two heuristic methods to identify SAT configuration of switches. Our first algorithm, coined WalkGrid, is an adaptation of the WalkSAT \cite{SelmanKautz94,ArdeliusAurell06}, which is a stochastic local search heuristic solver for the K-satisfiability problem. The WalkGrid algorithm is very fast and shows flawless performance in discovering a valid configuration almost anywhere in the SAT region.

Our second algorithm corresponds to solving BP equations. The BP scheme, designed in the spirit of \cite{02MPZ}, is used to find most biased/stressed link and then proceeds with decimation towards a valid configuration \cite{02MPZ}. Our implementation of the BP-decimation is so far slower and a bit less efficient than performance shown by the WalkGrid algorithm. To this point, let us note that there exists a more efficient way of using BP to find valid configurations - the reinforcement strategy \cite{ChavasFurtlehner05,BraunsteinZecchina06}, which is fully distributed, linear in the number of consumers and typically outperforms decimation. The reinforcement strategy has been implemented in \cite{09ABRZ} for a related online advertising problem. However, we had a difficulty to find an implementation of the reinforcement which would work efficiently in our problem.

\subsection{WalkGrid}
\label{sec:WalkGrid}

WalkGrid is a stochastic local search algorithm inspired and closely related to its K-SAT ancestor called WalkSAT \cite{SelmanKautz94,ArdeliusAurell06}. It can also be viewed as a Monte-Carlo-like algorithm which in order to gain speed violates the detailed balance condition (not needed here as we are not interested in sampling,  but are rather focused on a local search). Our implementation works as follows
\begin{codebox}
\Procname{$\proc{WalkGrid}$}
\li   Assign each value of $\sigma$ 0 or 1 randomly (but such that $\forall i \in G : \sum_{\alpha \in \partial i} \sigma_{i \alpha} =1 $);
\li  \Repeat   Pick a random power generator $\alpha$ which shows an overload, and denote the value of the overload, $\delta$;
\li   Choose a random consumer $i$ connected to the generator $\alpha$, i.e. $\sigma_{i\alpha}=1$; 
\li Pick an arbitrary other generator which is not overloaded and consider switching connection from $(i\alpha)$ to $(i\beta)$.
\li \If (in the result of this switch $\alpha$ is relieved from being overloaded 
\li {\bf and} $\beta$  either remains under the allowed load or it is overloaded but by the amount less than $\delta$)
\li Accept the move, i.e. disconnect  $i$ from $\alpha$ and connect it to $\beta$ thus setting $\sigma_{i\beta}=1$, $\sigma_{i\alpha}=0$.
\li \Else With probability $p$ connect consumer $i$ to $\beta$ instead of $\alpha$;
\li  \Until  Solution found or number of iterations exceeds $MT_{\rm max}$.
\end{codebox}

The WalkGrid algorithm depends on two parameters: the maximum number of iterations $T_{\max}$, and the temperature-like (greediness) parameter $p$. The parameter $p$ needs to be optimized, just as in the original WalkSAT solver.

\begin{figure}[!ht]
  \begin{center}
  \resizebox{0.48\linewidth}{!}{\includegraphics{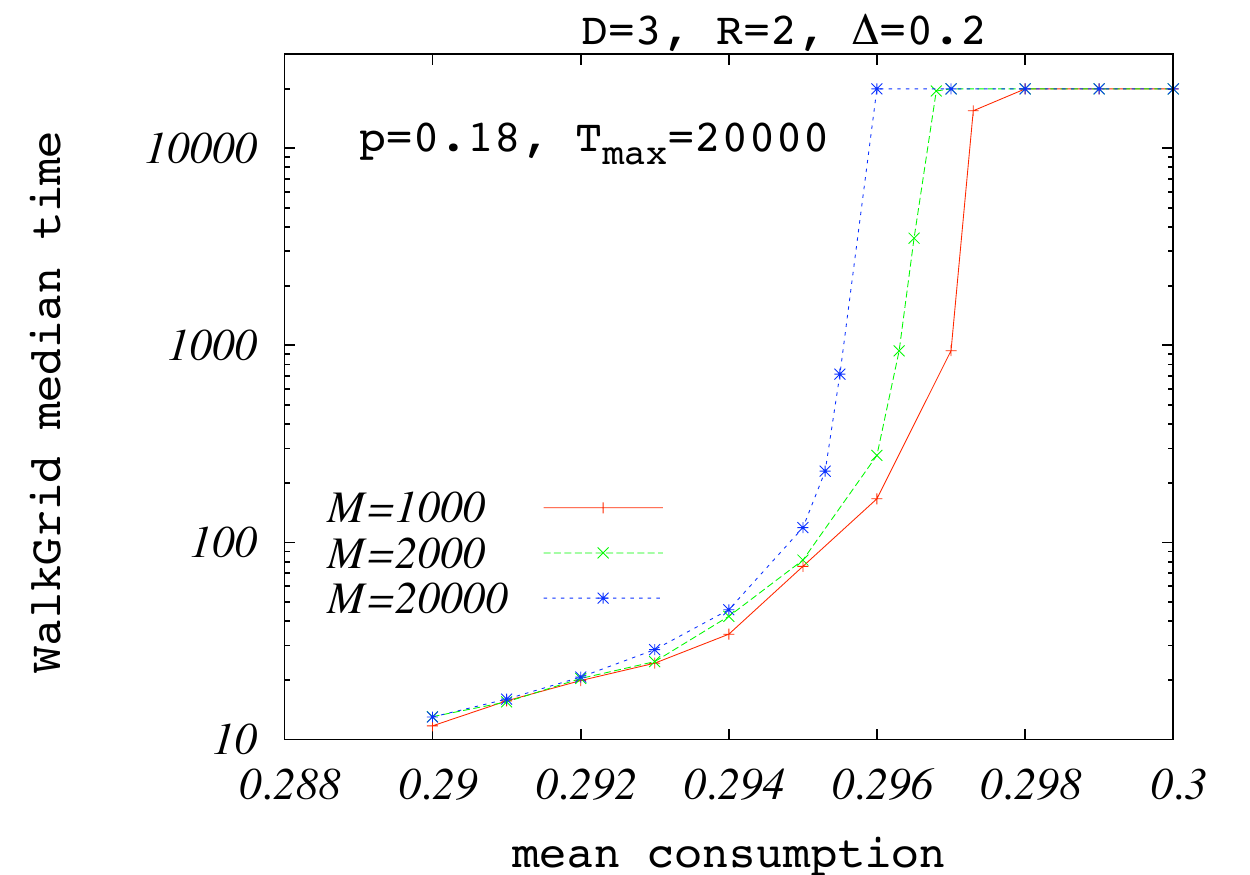}}
  \resizebox{0.48\linewidth}{!}{\includegraphics{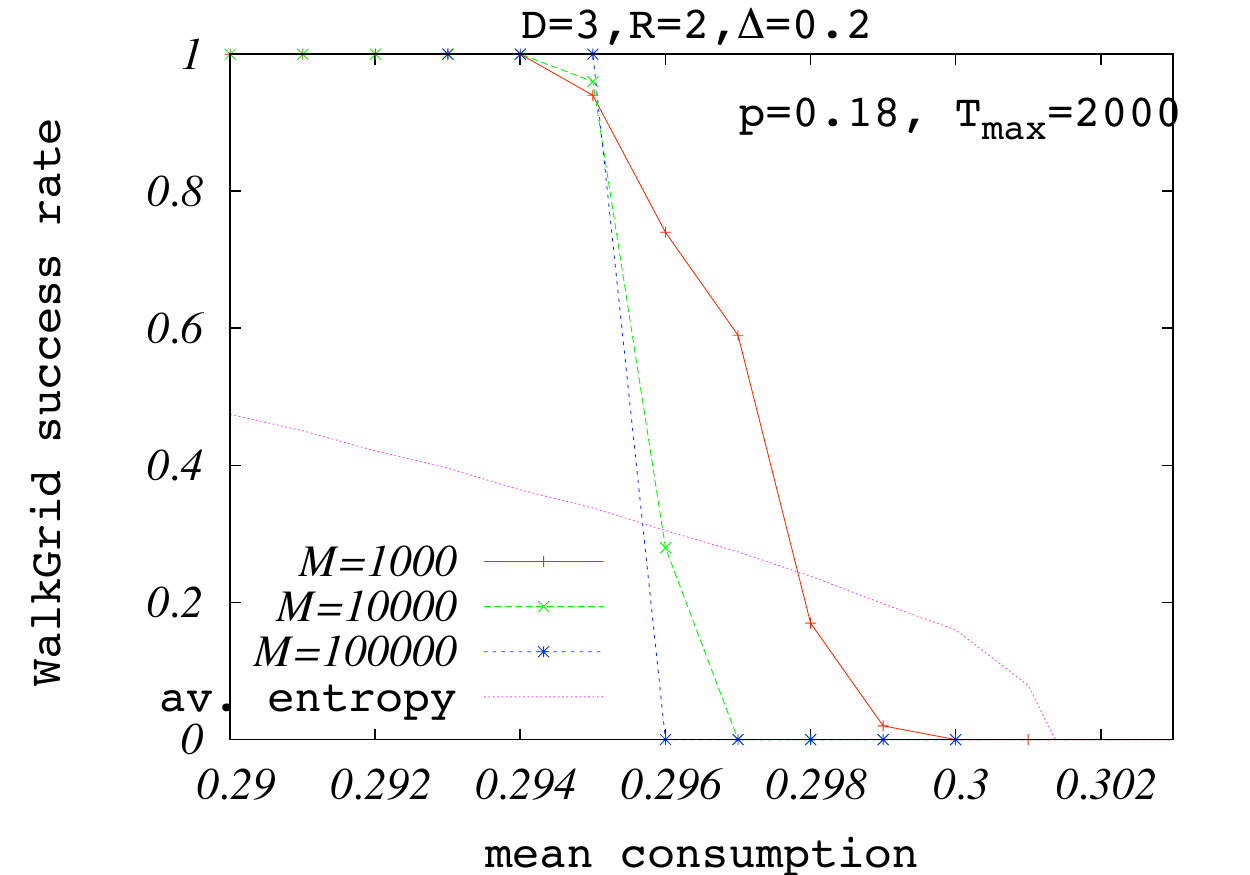}}
  \end{center}
  \caption{ (Color online) 
  Performance of the WalkGrid algorithm. The data are for networks with $M$ producers, and $R=2$, $D=3$, $\Delta=0.2$, $\epsilon=0$. The temperature-like parameter and the maximum number of iterations are set to $p=0.18$ and $T_{max}=2000$ respectively. The average (median) is over $100$ instances for $M$ up to $20k$, and $20$ instances for $M=100k$. Left: The median running time plotted against the average consumption $\overline x$. Right: The percentage of cases where solution was found in less than $T_{\rm max}=2000$ iteration. The Bethe entropy based (asymptotic) curve is drawn dashed for comparison (the actual value for the curve is not related to the success rate), suggesting that in the limit of $N\to \infty$ valid configurations exist up to $\overline x \approx 0.30$. Note that in the separated case (of $R=0$), valid configurations exist only at $\overline x\le 1/D - \Delta/2 \approx 0.23$. \label{fig_walkgrid}}
\end{figure}

Fig.~\ref{fig_walkgrid} shows performance of the WalkGrid algorithm. The running time of this algorithm scales close to linear with the system size, it is thus relatively easy to resolve fast network with many thousands of nodes. Whereas for $R=2$, $D=3$, $\Delta=0.2$ the separated architecture requires $\overline x \le 0.233$, the WalkGrid algorithm is able to find valid configurations up to $\overline x \approx 0.296$, while our theoretical analysis suggests that valid configurations should exist up to $\overline x \approx 0.301$.

\begin{figure}[!ht]
  \begin{center}
  \resizebox{0.45\linewidth}{!}{\includegraphics{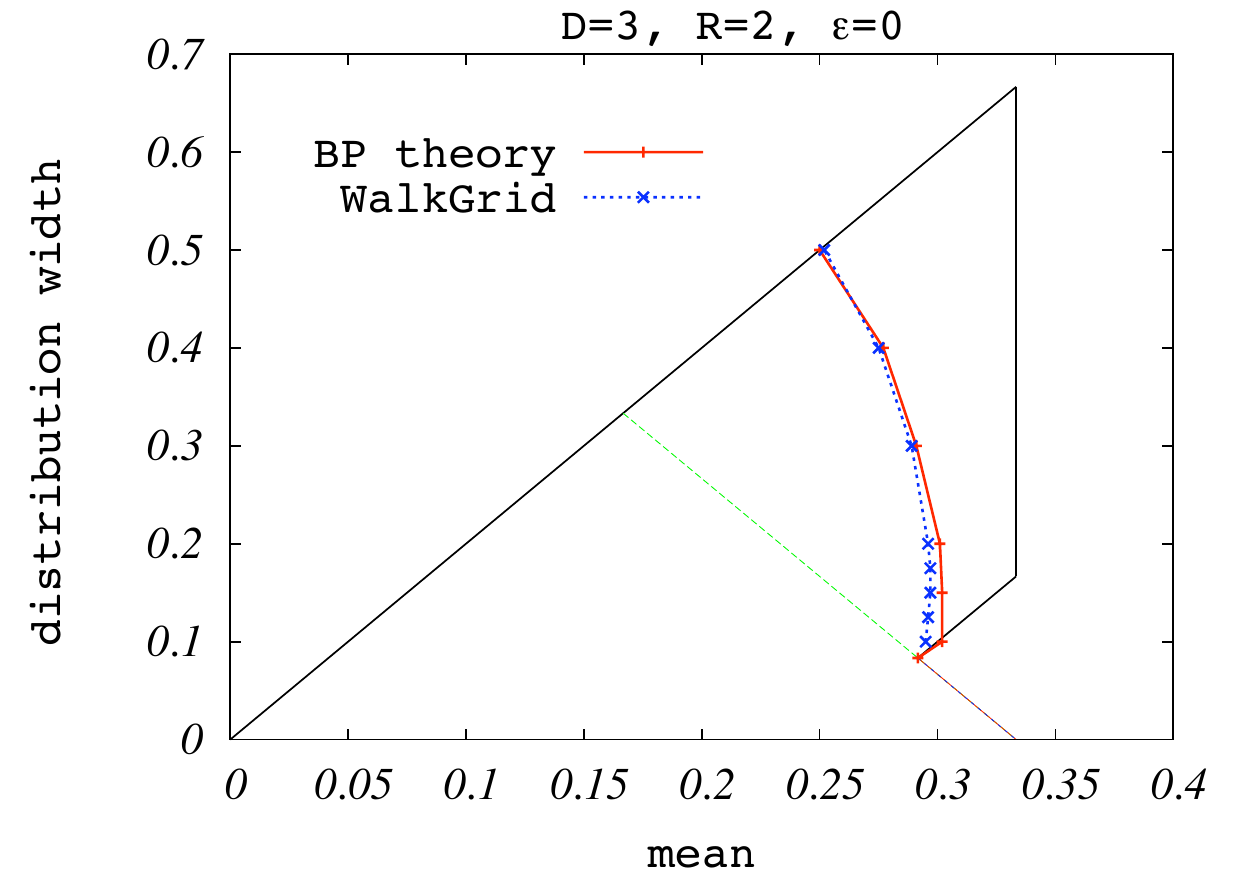}}
  \end{center}
  \caption{ (Color online)  
  \label{fig_WG_ph} Performance of the WalkGrid algorithm for $D=3$, $R=2$, $\epsilon=0$ and different values of the distribution width $\Delta$. Bellow the blue (dashed) line the WalkGrid algorithm is able to find solution on more than $50\%$ (average over $20$ trials) of graphs with $10^5$ generators. The red (full) line corresponds to the theoretical boundary separating SAT and UNSAT domains respectively.}
\end{figure}

\subsection{Belief-Propagation decimation}

In the BP-based decimation algorithm one updates Eqs.~(\ref{BP}-\ref{BP4}) iteratively, thus passing messages from generators to consumers and back. After fixed number of steps the most biased consumer is chosen and the more probable value for its consumption is assigned, the graph is reduced and the procedure is repeated.
Note that updating Eqs.~(\ref{BP}-\ref{BP2}) takes $2^{2R}$ steps per message thus making the algorithm exponential in $R$. However,
building new connections is expensive and one should realistically assume that actual $R$ is any case not very large. As far as the scaling in $N$ goes, the algorithm is quadratic in the number of consumers.

\begin{codebox}
\Procname{$\proc{Decimation}$}
\li \Repeat Update BP messages on every edge according to (\ref{BP}-\ref{BP4}) $n$ times.
\li         Compute the marginals (\ref{marg}).
\li         Choose the most biased edge $i\alpha$ and assign is the more probable values;
\li         Simplify the formula, cutting off the assigned edge from the graph;
\li \Until  Solution or contradiction is found;
\end{codebox}

\begin{figure}[!ht]
  \begin{center}
  \resizebox{0.45\linewidth}{!}{\includegraphics{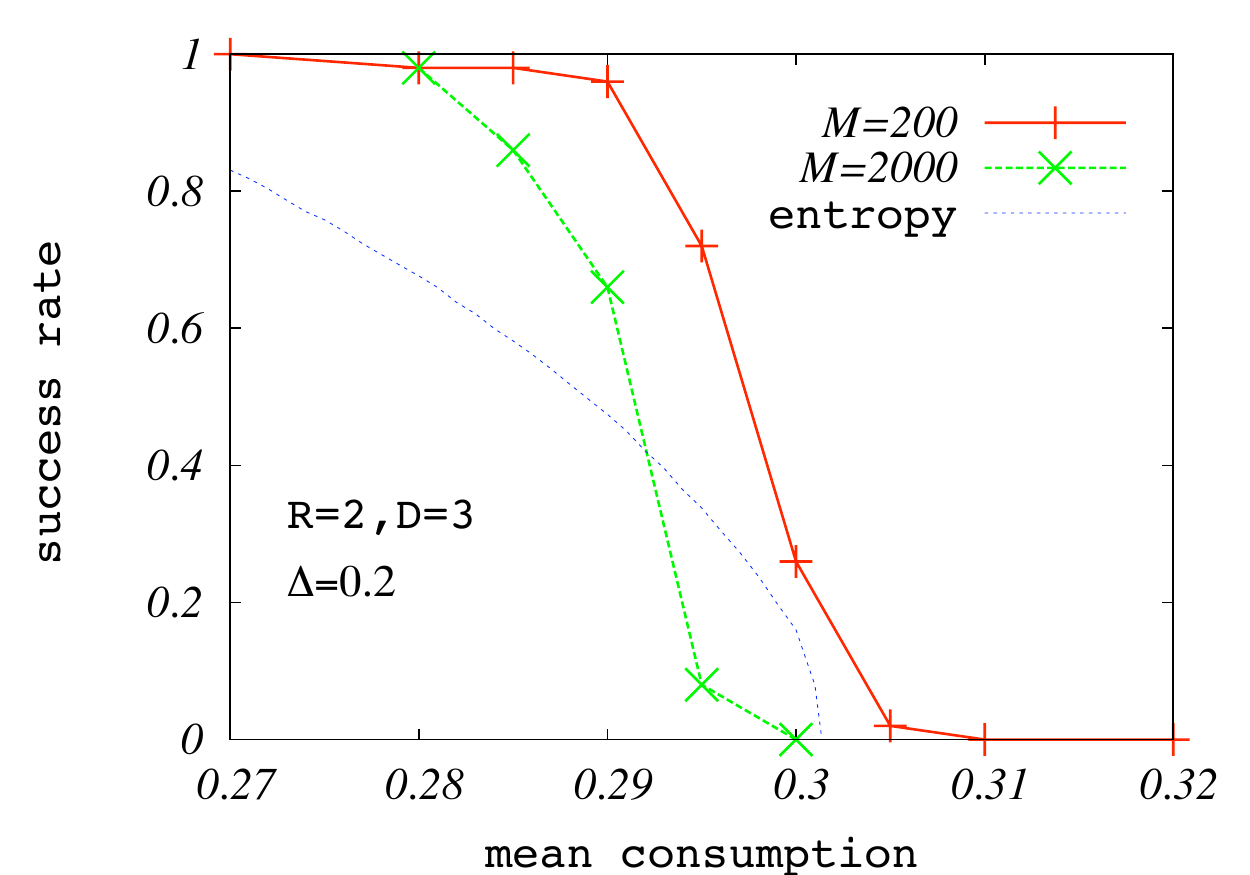}}
  \end{center}
  \caption{ (Color online)  \label{fig_dec} Performance of the BP-based decimation algorithm. The data are for networks with $M$ producers, $n=5$ (five iteration per a cycle of the decimation procedure) and $R=2$, $D=3$, $\Delta=0.2$, $\epsilon=0$. Average over $50$ random instances is taken and the fraction of successful runs is plotted against the mean consumption $\overline x$. The Bethe entropy based (asymptotic) curve is drawn dashed for comparison (the actual value for the curve is not related to the success rate), suggesting that in the limit of $N\to \infty$ valid configurations exist up to $\overline x \approx 0.30$. Note that in the separated case (of $R=0$), valid configurations exist only at $\overline x\le 1/D - \Delta/2 \approx 0.23$.}
\end{figure}

The algorithm performance is illustrated in Fig.~\ref{fig_dec} where the percentage of success in the BP-based decimation is shown. These data  average over $50$ random instances from the $R=2$, $D=3$ ensemble with variance in consumption $\Delta=0.2$ and varying mean consumption. For such a set of parameters the separated network would be able to achieve $\overline x \le 1/3-0.1$, whereas with two consumers per producer connected to two producers the mean consumption increases to $\overline x \approx 0.30$.

\section{Summary and Path Forward}
\label{sec:Con}

This manuscript reports a first study of the power distribution networks with transition from the SAT regime, in which shedding of loads is avoidable, to the UNSAT regime, where shedding is the only available option for balancing the demand. We have shown that a significant enlargement of the SAT-domain is possible employing ancillary connections between power consumers and generators. Even though our model represents a gross oversimplification over the actual power grid, it offers a significant step forward in providing a framework and guidance for analysis of more involved and realistic problems. The general approach we pursued in this study is based on recent developments in the field of graphical models that merges statistical physics,  computer science, optimization theory and information theory \cite{08RU,09MM}.
The BP approach is asymptotically exact on infinite sparse graphs and as such is useful for the asymptotic (capacity/phase-transition style) analysis.
The BP scheme also provides heuristic tools for graphical models on finite sparse graphs that can be used for algorithmic optimization and control of the power grid. We tested this BP approach, and also developed in parallel another an apparently more efficient alternative to BP called WalkGrid. This algorithm finds valid configuration of switching practically anywhere inside the SAT phase. Note,  that the two algorithms, BP and WalkGrid, are truly complementary and there utility for practical problems in power networks are yet to be explored.

It is important to emphasize that many generalizations of our model are very straightforward and can be used directly within the framework presented here. This includes implementing different probability distributions for demands (as long as the distribution support is bounded), and non-uniformity for both generator and consumer levels, i.e. varying production caps for generators and introducing distinct distributions of demands for different consumers.  Also the network itself can be easily extended in many ways from the simplified case, parameterized by $R$ and $D$, that we discussed in the manuscript. Our equations are straightforwardly valid for every bipartite locally tree-like random network. Another case, which allows very natural and straightforward generalization for all the statements made in this manuscript, corresponds to breaking the equivalence between different edges in the graph and thus assigning nonuniform weights to them.  These weights may e.g. represent cost of construction, geographical length, proxy for losses, cost of exploitation, etc.

There are many other more realistic extensions of our model associated with the description, optimization and control of power grids which can  benefit from utilizing a graphical model approach of the kind discussed in this paper,  even though actual implementation may prove to be more involved. We conclude listing some of these more interesting but difficult problems that we plan to address in the future, based on the general method  sketched in this manuscript:
\begin{itemize}
\item [A.] Most important generalization of our approach would be to account for losses, impedances and the reactive character of AC electrical systems. Obviously this will require incorporating in our statistical SAT-UNSAT framework Kirchhoff's circuit laws \cite{96WW,98Bol}.

\item [B.] Generators in a real grid are interconnected on higher (still power distribution, but also power generation) levels. In combination with item [A], this represents a major challenge for extending our approach. However,  we still believe that posing the joint optimization and control problem in terms of a complex graphical model and then addressing capacity/transition as well as algorithmic issues with the host of BP-related techniques is a feasible and exciting path forward.

\item [C.] The joint optimization setting, mentioned in item [B], may also include various additional factors associated with
economic policies (e.g. prices, incentives, etc) \cite{01BH,08WBAF}, government regulations and load/generation forecasting. These complications can be accounted for in the form of extra soft or hard constraints. This united framework should also take advantage of the progress made in developing optimal power flow solutions \cite{91HG,08PJ}.

\item [D.] Our consideration in this paper was purely static,  thus ignoring important transients. Constructing dynamical (discrete and continuous time) models that also accounts for all the aforementioned problems, is an important future task. It is also important to notice that the graphical model approach can in fact be extended to the dynamic framework, see e.g. \cite{04HND},  thus suggesting yet another intriguing future opportunity.

\item [E.] An essential part of statistical studies of the power grid focuses on estimating probabilities and mitigating very costly and dangerous large scale outages \cite{07DCLN,08HT}. The graphical model framework this manuscript describes allows an extension which can analyze rare events of special interest, such as dangerous but rare configurations of the channel noise and degrading performance of Low-Density-Parity-Check codes \cite{04CCSV,08CS}. We plan to extend this approach to the analysis of outages and their cascading through the grid.

\end{itemize}

\section{Acknowledgments}

We thank Florent Krzakala, David Gamarnik and Scott Backhaus for very stimulating and critical comments, and all members of LANL study group on "Optimization and Control Theory for Smart Grids" for many discussions. The work at LANL was carried out under the auspices of the National Nuclear Security Administration of the U.S. Department of Energy at Los Alamos National Laboratory under Contract No. DE-AC52-06NA25396.

\bibliographystyle{apsrev}
\bibliography{GridRed}

\end{document}